\documentclass[
reprint,
amsmath,amssymb,
aps,
prb,
floatfix,
]{revtex4-2}

\usepackage{graphicx}
\usepackage{dcolumn}
\usepackage{bm}
\usepackage{tikz}
\usepackage{ bbold }

\newcommand{\half}{\frac{1}{2}}
\newcommand{\sech}{\text{sech}}

\begin{document}

\preprint{APS/123-QED}

\title{Thermal Spin Waves from Accelerating Domain Walls via the Unruh Effect}

\author{A.L. Bassant$^1$}
 \email{a.l.bassant@uu.nl.}
\author{R.A. Duine$^{1,2}$}
\affiliation{$^1$Institute for Theoretical Physics, Utrecht University, Princetonplein 5, 3584CC Utrecht, The Netherlands \\
$^2$Department of Applied Physics, Eindhoven University of Technology, P.O. Box 513, 5600 MB Eindhoven, The Netherlands}

\author{}
\affiliation{}

\date{\today}

\begin{abstract}
We consider a wire consisting of a conducting ferromagnetic layer and an insulating antiferromagnetic layer that are coupled.
The ferromagnet hosts a domain wall, which is dynamically driven by a charge current. 
We show that for a specific time-dependent current, the domain wall moves according to a Rindler trajectory.
This motion excites spin waves in the antiferromagnetic insulator, and their emission spectrum is characterised by an effective temperature analogous to the Unruh temperature, $T_U=\hbar a / 2\pi c k_B$ with $a$ the acceleration of the domain wall, $c$ the maximum antiferromagnetic spin wave velocity, and $k_B$ the Boltzmann constant. 
This thermal signature is a direct consequence of the Unruh effect and could be experimentally observed. 
Our results establish magnetism as a promising platform for probing relativistic quantum field phenomena.
Moreover, since the Unruh effect is inherently linked to entanglement, our proposal provides a route for entangling magnetic domain walls via relativistic effects.
\end{abstract}

\maketitle

\section{Introduction}

In 1973, Fulling \cite{fulling_nonuniqueness_1973}, Davies \cite{davies_scalar_1975}, and Unruh \cite{unruh_notes_1976} predicted that an accelerated observer would see an ambient temperature arising from the vacuum.
This is known as the Fulling-Davies-Unruh effect or simply the Unruh effect.
This prediction has not been observed in actual spacetime since the acceleration necessary to find 1K of radiation is $2.47\times 10^{20}$m/s$^2$, which is approximately 9 orders of magnitude larger than the surface gravity of a neutron star \cite{NeutronStar}.
Despite these exceptionally large accelerations, some attempts have been made using circular acceleration \cite{rad_test_2012}.
Other efforts aim to experimentally demonstrate the effect using a Bose-Einstein condensate or graphene, where relativistic quantum mechanics manifests in an effective manner.

These condensed matter analogues of space-time allow for the detection of the Unruh effect, since the necessary acceleration is significantly reduced.
The reduction is due to the analogue of the speed of light being much smaller than the actual speed of light.
For a Bose-Einstein condensate, this is the speed of sound, and for graphene, this is the Fermi velocity.
Some theoretical proposals consist of manipulating these velocities to create a non-inertial observer \cite{dv:bhardwaj_unruh_2023,dv:tallent_analog_2024} or implementing circular acceleration in a Bose-Einstein condensate \cite{ca:biermann_unruh_2020,ca:gooding_interferometric_2020,ca:d_bunney_circular_2023}.
Others propose to quench the system with an internal drift velocity to simulate the change in inertial observer \cite{Quench:kosior_unruh_2018,Quench:rodriguez-laguna_synthetic_2017,Quench:hu_quantum_2019}.
Many of these proposals concern a circular path or quench rather than a linear setup.
Nonetheless, linear setups have been considered in the following articles:
in Ref. \cite{la:retzker_methods_2008}, they theoretically investigate accelerating atom dots in a Bose-Einstein condensate, where the atom dots act as detectors of the Unruh effect.
Similarly, Ref. \cite{sc:blencowe_analogue_2020} uses superconductivity with microwave circuits that are rows of Josephson junctions and bulk acoustic resonators to make a moving event horizon.
These theoretical proposals all rely on the analogue gravity effects of condensed-matter setups. 
However, experimental evidence for analogue gravity has been observed only once in Ref. \cite{BEC_BH}, as the necessary conditions for these effects are still difficult to achieve. 

To address this challenge, a new class of analogue gravity setups has recently been proposed: magnetic systems. 
These systems are expected to be versatile, because many properties can be controlled to a large extent.
As a result, these could prove to be valuable for analogue gravity experiments.
In fact, theoretical studies have already demonstrated their potential, showing that magnetic analogues can simulate phenomena such as Hawking radiation in a magnonic black and white hole \cite{roldan-molina_magnonic_2017}, and superradiance in the context of the magnetic Klein paradox \cite{bassant_entangled_2024,harms_enhanced_2022,errani_negative-energy_2025}.
The emergent relativistic theory in antiferromagnets has also been extensively investigated for antiferromagnetic domain walls \cite{AFMDW:alliati2022relativistic,AFMDW:shiino2016antiferromagnetic,AFMDW:yang2019atomic,tatara_magnon_2020,SWinAFM}.
In particular, excitations on top of the antiferromagnetic ordering—known as spin waves—are described by a relativistic field theory in the long-wavelength limit. 

Leveraging the tunability of magnetic systems, we propose a magnetic setup that explicitly demonstrates an analogue of the Unruh effect.
The setup consists of a wire with a conducting ferromagnetic layer and an insulating antiferromagnetic layer that are coupled via a nonmagnetic spacer layer, as shown in Fig. \ref{fig:setup}. 
We consider the case that the magnetisation of the ferromagnet features a domain wall, which is moved using a charge current in the wire.
The charge current varies in time such that the domain wall moves according to a Rindler trajectory.
This is a trajectory of constant acceleration in a relativistic theory defined by the long-wavelength limit of the spin waves in the antiferromagnet \cite{Rindler}.
The accelerating domain wall emits spin waves in the antiferromagnet, and we theoretically demonstrate that the emission is characterised by the Unruh temperature, $$T_U=\frac{\hbar a }{ 2\pi c k_B},$$ where $a$ is the acceleration of the domain wall, $c$ is the maximum spin wave velocity, and $k_B$ is the Boltzmann constant.
This is a direct consequence of the Unruh effect, which could become observable with modest improvements to current experimental techniques. 
Consequently, magnetic setups would not only enable the observation of the Unruh effect but also render other analogue gravity phenomena experimentally accessible. 
These advantages position magnetic systems as promising candidates for future research into quantum simulators of relativistic field theories.

Besides this fundamental interest, we may be able to leverage entanglement that arises in relativistic theories.
For example, it has been shown that two co-accelerating two-level atoms become entangled because of the Unruh effect \cite{benatti_entanglement_2004}.
By analogy, two co-accelerating domain walls may also become entangled.
Furthermore, the analogue Unruh effect is independent of the specific magnetic texture in the ferromagnet.
Consequently, any movable topological texture—such as a skyrmion—could be employed to observe similar phenomena.
Therefore, this approach offers a novel strategy for entangling magnetic textures, which benefits the emerging field of quantum magnonics \cite{Yuan_Cao_Kamra_Duine_Yan_2022}, an interdisciplinary line of research that aims to leverage the intersection of magnetic dynamics and quantum information for innovative information processing.

The rest of the article is structured as follows: in Sec. \ref{sec:setup and Model}, we explain the theoretical setup and derive the equation of motion for spin waves in the antiferromagnet.
In Sec. \ref{sec:Rindler trajectory}, we provide the necessary conditions for the ferromagnetic domain wall to accelerate according to a Rindler trajectory.
Sec. \ref{sec:The magnetic FDU effect} shows how the Unruh effect manifests in the emission of antiferromagnetic spin waves from the accelerating domain wall. 
In addition, we also compute the average spin wave amplitude for different accelerating domain walls.
Finally, we conclude and discuss in Sec. \ref{sec:Conclusion}.

\section{Setup and Model}\label{sec:setup and Model}

\begin{figure}
    \centering
   \includegraphics[width=\linewidth]{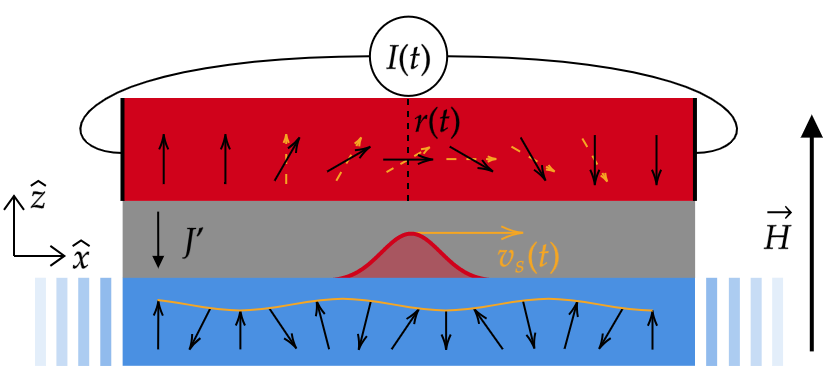}
    \caption{This is a schematic of a magnetic wire in a magnetic field $\vec H$. The wire consists of a blue layer, which is an insulating antiferromagnet, and the grey in-between layer is the nonmagnetic spacer that provides RKKY interaction with strength $J'$. The red layer is a conducting ferromagnet featuring a domain wall at position $r(t)$. The domain wall is a source for spin waves in the antiferromagnet, which is represented by the red Gaussian. The orange attributes represent the time-dependent dynamics: a time-dependent current $I(t)$ in the ferromagnet drives domain wall movement, as indicated by the dashed orange arrows. This, in turn, displaces the source in the antiferromagnet, which is depicted by the arrow with a time-dependent velocity, $v_s(t)$.
    The moving domain wall causes spin waves to be excited in the antiferromagnet.}
    \label{fig:setup}
\end{figure}

The setup is shown in Fig. \ref{fig:setup}, which consists of a conducting ferromagnetic layer and an insulating antiferromagnetic layer that are weakly coupled by an exchange interaction with strength $J'$.
The weak coupling is achieved by a nonmagnetic layer between the ferromagnetic and antiferromagnetic layers, which allows for Ruderman-Kittel-Kasuya-Yosida (RKKY) interaction \cite{RKKY1,RKKY2}.
In this section, we show that a domain wall in the ferromagnetic layer of the wire manifests as a source for spin waves in the antiferromagnetic layer.
In this discussion, we assume that the antiferromagnet has negligible damping, since it is an insulator.

We consider a two-sublattice antiferromagnet with uni-axial easy-axis anisotropy, which we take along the $z$-direction.
This allows us to write the antiferromagnetic Lagrangian in terms of the total magnetisation in a unit cell ($\vec m =\vec m_1+\vec m_2$ with $\vec m_i$ the unitless magnetisation of the sublattice) and the Ne\'el vector of the unit cell, $\vec n=(\vec m_1-\vec m_2)/2$.
The Lagrangian is therefore given by, \cite{AFM1,AFM2,SWinAFM}

\begin{align}
    \mathcal{L}[\vec n,\vec m]=&\frac{M_s}{\gamma}\vec m\cdot(\dot{\vec n}\times \vec n)+M_s\vec m\cdot (\vec H +J'\vec n_{FM}) \\
    &-\half J |\vec m|^2-\half(A|\partial_x\vec n|^2-K_zn_z^2), \nonumber
\end{align}

\noindent where $M_s$ is the saturation magnetisation, $\gamma$ is the gyromagnetic ratio, $\vec H=H\hat z$ is the external magnetic field, $J'$ is the RKKY interaction, $\vec n_{FM}$ is the magnetisation of ferromagnetic layer, $J$ is the intersublattice exchange, $A$ is the spin stiffness, and $K_z$ is the easy-axis anisotropy along the $z$-axis.
The dynamics of $\vec m$ depend only on $\vec n$, which makes it a slave variable.
Therefore, we integrate $\vec m$ out.
This leaves us with the effective Lagrangian for the Ne\'el vector,

\begin{align}
    \mathcal{L}[\vec n]=&\frac{M_s^2}{2J\gamma^2}\big|\dot{\vec n}-\gamma (\vec H+J'\vec n_{FM})\times \vec n\big|^2 \\
    &-\half(A|\partial_x\vec n|^2-K_zn_z^2). \nonumber
\end{align}

\noindent Linearising the Lagrangian using $\vec n=\hat z+\delta \vec n$ and $\delta \vec n=(\delta n_x,\delta n_y,\half\delta n_x^2 + \half\delta n_y^2)$ gives us the Lagrangian for antiferromagnetic spin waves in the long wavelength limit.
The spin wave Lagrangian is derived in Appendix \ref{Ap:Li} and it is given by,

\begin{align}
    \mathcal{L}[\vec n]=&\half\varphi^*\left(-\frac{M_s^2}{2J\gamma^2}(\partial_t+i\gamma H)^2+A\partial_x^2+K_z\right)\varphi \\
    &-\half A j^* \varphi-\half A j\varphi^*. \nonumber
\end{align}

\noindent Where $\varphi=\delta n_x+i\delta n_y$ and the source is given by $j=M_s^2(\gamma^2\eta H+i\gamma \dot \eta)/JA\gamma^2$ with $\eta= J'(\vec n_{FM}^x+i\vec n_{FM}^y)$.
In the derivation, we used that $|\vec H|\gg 2|J'\vec n_{FM}|$ and that the quadratic terms in $J'\vec n_{FM}$ vanish since the coupling $J'$ is weak.
The magnetic field is a constant shift of the frequency, which is absorbed into the time derivative.
Varying this Lagrangian gives us the equation of motion for the complex field,

\begin{equation}\label{eq:eom1}
    \frac{M_s^2}{J\gamma^2}\partial_t^2\varphi-A\partial_x^2\varphi+K_z\varphi=A j(x,t).
\end{equation}

\noindent The dispersion relation that follows from the equation of motion is $\omega(k)=\sqrt{J\gamma^2(A k^2+K_z)}/M_s$.
Herein, the analogue to the speed of light, the maximum spin wave velocity, is $c=\sqrt{J\gamma^2A}/M_s$.
The obtained equation of motion for antiferromagnetic spin waves has a source term, which is a direct consequence of the ferromagnetic dynamics.
Later on, we use a ferromagnetic domain wall as a source for antiferromagnetic spin waves.

We have just shown that the source is a consequence of the ferromagnetic dynamics, which are governed by the ferromagnetic Lagrangian,

\begin{align*}
    \mathcal{L}_{FM}=&\hbar (\cos(\theta(x,t))-1)\left(\partial_t+v_s\partial_x\right)\phi(x,t) \\
    &-\frac{J_{FM}}{2}\left((\nabla \theta(x,t)+\sin^2(\theta(x,t))(\nabla \phi(x,t))^2\right)\\
    &-H \cos(\theta(x,t))+\frac{K_\perp^{FM}}{2}\sin^2(\theta(x,t))\sin^2(\phi(x,t)) \\
    &-\frac{K_z^{FM}}{2}\cos^2(\phi(x,t)).
\end{align*}

\noindent The angles $\theta(x,t)$ and $\phi(x,t)$ are given by the parametrisation of the magnetisation of the ferromagnet, $\vec n_{FM}=(\cos\phi\sin\theta,\sin\phi\sin\theta,\cos\theta)$.
The first term corresponds to the Berry phase with a charge current component $v_s$, as shown in Eq. (256) of \cite{TATARA2008213}.
The other terms are exchange ($J_{FM}$), magnetic field ($H$), and anisotropy ($K_{\perp}^{FM}, K_z^{FM}$).
Additionally, we take damping effects into account, since the ferromagnet is a conductor, and it is essential for domain wall motion.
The damping is given by the Rayleigh dissipation functional,

\begin{align*}
    \mathcal{R}=\frac{\hbar \alpha_G}{2}\left|\left(\partial_t+\frac{\beta}{\alpha_G}v_s\partial_x\right)\vec n_{FM}\right|^2.
\end{align*}

\noindent Here, the damping is parametrised by the Gilbert damping $\alpha_G$.
The charge current also enters here, where the dissipative spin transfer torque is characterised by $\beta$.
In the next section, we use the ferromagnetic Lagrangian and Rayleigh dissipation functional to determine the equations of motion of the domain wall.
These equations allow us to derive the time-dependent charge current that is necessary to move the domain wall along a Rindler trajectory.
In the section thereafter, we use this result together with Eq. (\ref{eq:eom1}) to consider the magnetic analogue of the Unruh effect.

\section{Rindler trajectory for a ferromagnetic domain wall}\label{sec:Rindler trajectory}

In this section, we determine the time-dependent charge current required for a domain wall to move along a Rindler trajectory (for a review on current-driven domain wall dynamics, see Ref. \cite{TATARA2008213}). 
To achieve this, we derive the domain wall Lagrangian from the ferromagnetic Lagrangian.
This follows from substituting the domain wall ansatz in the Lagrangian of the ferromagnet and integrating over the spatial variables.
The domain wall ansatz is given by $\theta(x,t)=2\arctan(\exp(-(r(t)-x)/\lambda))$ with $\lambda=\sqrt{J_{FM}/K_z^{FM}}$ , and we assume that $\phi(x,t)=\phi_0(t)$ is spatially independent.
Therefore, the dynamics of the domain wall are described by the domain wall position $r(t)$ and an internal angle $\phi_0(t)$.
This functional form of the domain wall is accurate for thin wires, such that the magnetisation is homogeneous in the $y$ and $z$ components.
After substituting the ansatz and integrating over the spatial variables of the ferromagnetic Lagrangian, we find,

\begin{align}
    L=&N\bigg(\frac{\dot r(t)-v_s(t)}{\lambda}\dot \phi_0(t)-\frac{K_{\perp}^{FM}}{2}\sin^2\phi_0(t)+ H \frac{r(t)}{\lambda}\bigg).\nonumber
\end{align}

\noindent The prefactor $N$ is given by $2\lambda L_yL_z/a^3$ where $L_y$ and $L_z$ are the dimensions of the wire in the $y$ and $z$ direction.
The lattice spacing is given by $a$.
Similarly, we compute the Rayleigh dissipation functional for the domain wall.

\begin{align*}
    R=\frac{\hbar \alpha_G N}{2}\left(\frac{1}{\lambda^2}\left(\frac{\beta}{\alpha_G}v_s-\dot r(t)\right)^2+(\dot \phi_0(t))^2\right).
\end{align*}

\noindent From the domain wall Lagrangian and the Rayleigh dissipation functional, the equations of motion of the domain wall are derived.
They are given by \cite{DW:Tatara,TATARA2008213},

\begin{subequations}
\begin{align}
    \dot\phi_0(t)+\alpha_G\frac{\dot r(t)}{\lambda}=&H+\frac{\beta}{\lambda}v_s(t),\\
    \frac{\dot r(t)}{\lambda}-\alpha_G\dot\phi_0(t)=&\frac{K_\perp^{FM}}{2\hbar}\sin 2\phi_0(t)+\frac{v_s(t)}{\lambda}.
\end{align}
\end{subequations}

\noindent These equations allow us to isolate $v_s(t)$:

\begin{subequations}
\begin{align} \label{eq:vs}
    (1+\alpha_G\beta)\frac{v_s(t)}{\lambda}=&(1+\alpha_G^2)\frac{\dot r(t)}{\lambda}-\frac{K_\perp^{FM}}{2\hbar}\sin2\phi_0(t)-\alpha_G H,\\
    (1+\alpha_G \beta)\dot\phi_0(t)=& H+(\beta-\alpha_G)\frac{\dot r(t)}{\lambda}- \beta\frac{K_\perp^{FM}}{2\hbar}\sin 2\phi_0(t). \label{eq:phi}
\end{align}
\end{subequations}

\noindent Solving the right-hand side of Eq. (\ref{eq:vs}) gives us the time-dependent charge current that is required to get a specific $r(t)$.

Now, we compute the time-dependent charge current for the accelerating domain wall.
The ferromagnetic domain wall follows a path of constant acceleration from the point of view of the relativistic dynamics that is set by the spin wave velocity $c$ of the antiferromagnet.
This is the so-called Rindler trajectory, $r_{\text{rind}}(t)=c\sqrt{t^2+c^2e^{2a \xi/c^2}/a^2}$.
This hyperbolic path is parametrised by the acceleration $a$ and a spatial coordinate $\xi$.
The spatial coordinate $\xi$ is determined by the initial position of the domain wall, $x_0$, according to the relation $\xi=c^2\log(a x_0/c^2)/a$.
The right-hand side of Eq. (\ref{eq:vs}) also depends on the internal angle, which follows from Eq. (\ref{eq:phi}).
This equation simplifies when $|H -(\alpha_G-\beta)\text{Max}(v_s(t))/\lambda|<|\alpha_G K_\perp^{FM}/2\hbar|$.
Then we have that $\dot\phi_0(t)=0$, which is known as the Walker breakdown \cite{WalkerBreakdown}. 
If the criterion is not satisfied, then one has to resort to numerical means.
We assume that this inequality is satisfied, i.e. that we are below the Walker breakdown.
Therefore, we simplify Eq. (\ref{eq:vs}) to find,

\begin{align} \label{eq:vsanalytical}
    v_s(t)=&\frac{\alpha_G}{\beta}\dot r_{\text{rind}}(t)-\frac{\lambda}{\beta}H.
\end{align}

\noindent This time-dependent charge current causes the domain wall to move according to a Rindler trajectory.

\begin{figure}
    \centering
    \includegraphics[width=\linewidth]{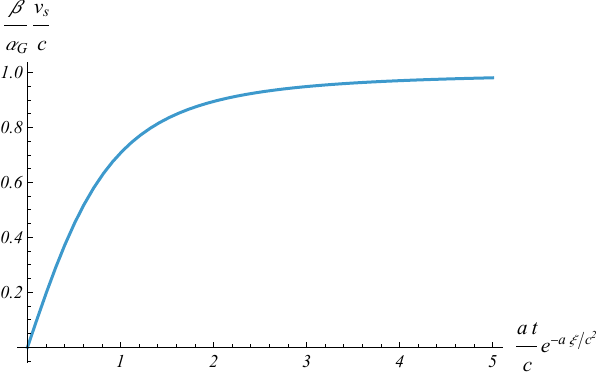}
    \caption{This is the time-dependent current necessary to accelerate the domain wall according to a Rindler trajectory, which is given by Eq. (\ref{eq:vsanalytical}). We have taken $H=0$, since taking nonzero $H$ simply shifts the graph.}
    \label{fig:CurrentProfile}
\end{figure}

The charge current initially grows linearly in time and then converges to its maximum value, given by $(\alpha_G c-\lambda H)/\beta$, see Fig. \ref{fig:CurrentProfile}.
Interestingly, the magnetic field strength can be tuned in such a way that the maximum current is experimentally achievable.
In addition, the current must be below the Walker breakdown, $| H -(\alpha_G-\beta)c/\lambda|<|\beta K_\perp^{FM}/2\hbar|$, for this solution to hold, and it should not exceed the critical current for domain wall nucleation \cite{DWcurrentcondition}.

\section{The Unruh effect}\label{sec:The magnetic FDU effect}

In the previous section, we established the time-dependent current for which the ferromagnetic domain wall follows a Rindler trajectory as seen from the spin waves in the antiferromagnet.
In this section, we compute the average spin wave amplitude generated from the moving domain wall.
First, we solve Eq. (\ref{eq:eom1}) for a stationary domain wall and define the average spin wave amplitude using stationary modes. 
Next, we solve Eq. (\ref{eq:eom1}) for a domain wall moving along a Rindler trajectory, yielding the spin wave emission in terms of accelerated modes.
These accelerated modes are then transformed into a basis of stationary modes, which allows for the evaluation of the average spin wave amplitude associated with the moving domain wall.
Our analysis reveals that the moving domain wall excites spin waves with an average amplitude that is characterized by the Unruh temperature.

We recast Eq. (\ref{eq:eom1}) into a simpler form,

\begin{equation}\label{eq:eom2}
     \frac{1}{c^2}\partial_t^2\varphi-\partial_x^2\varphi+\mu^2\varphi=j(x,t).
\end{equation}

\noindent Where we define $\mu=\sqrt{K_z/A}$.
This is known as the Klein-Gordon equation, which describes a relativistic scalar field.
The homogeneous solutions of Eq. (\ref{eq:eom2}) satisfy the dispersion relation $\omega(k)=c\sqrt{k^2+\mu^2}$.
These solutions are called stationary modes.
The general solution to Eq. (\ref{eq:eom2}) is then given by,

\begin{equation}
    \varphi(x,t)=\int dk d\omega_0\frac{c^2 J_M(k,\omega_0)e^{i kx-i\omega_0t}}{c^2(k^2+\mu^2)-\omega_0^2-i\varepsilon\omega_0},
\end{equation}

\noindent where $\varepsilon$ is a small imaginary offset that makes the integral calculable and $J_M(k,\omega_0)$ is the source of spin waves represented in stationary modes with wavevector $k$ and frequency $\omega_0$,

\begin{equation}\label{eq:Jm}
    J_M(k,\omega_0)=\int dxdt \hspace{3px} j(x,t)e^{-i k x+i\omega_0t}.
\end{equation}

\noindent The general solution is further simplified through contour integration, resulting in

\begin{align}\label{eq:solM}
    \varphi(x,t)= c^2\pi i\int  \frac{dk}{\omega(k)}&\big[J_M(k,\omega(k))e^{i k x-i\omega(k) t} \\
    &-J_M(k,-\omega(k))e^{i k x+i\omega(k) t}\big].\nonumber
\end{align}

\noindent This expression gives the emission of spin waves due to the domain wall.

The emitted spin waves cause the antiferromagnet to slightly deviate from the equilibrium N\'eel vector.
This is quantified by the spatially and temporally averaged amplitude of the spin waves, which is defined as

\begin{equation}
    A_M=\int \frac{dxdt}{VT}\hspace{3px} |\varphi(x,t)|^2\approx\int \frac{dxdt}{VT}\hspace{3px}|\hat z -\vec n|^2.
\end{equation}

\noindent The averaging is performed over volume $V$ and time $T$.
The quantity $A_M$ will be used to characterise the Unruh effect.
First, we determine the relation between $J_M$ and the average spin wave amplitude using the solution of Eq. (\ref{eq:solM}),

\begin{align}\label{eq:final}
    A_M =\frac{ c^4\pi^2}{V} \int   \frac{dk }{\omega^2(k)}&(|J_M(k,\omega(k))|^2 \\
    &+|J_M(k,-\omega(k))|^2).\nonumber
\end{align}

Now, we solve Eq. (\ref{eq:eom2}) with a domain wall that follows a Rindler trajectory.
The coordinates of a Rindler trajectory are given by, \cite{higuchi_fulling-davies-unruh_1993,Rindler}

\begin{align}\label{eq:Rtrans}
    t=& \frac{c}{a}e^{a\xi/c^2}\sinh{\frac{a\tau}{c}}, & r_{\text{rind}}=&\frac{c^2}{a}e^{a\xi/c^2}\cosh{\frac{a\tau}{c}}.
\end{align}

\noindent Where $a$ is the proper acceleration and $r_{\text{rind}}$ is written in terms of proper time \footnote{The definition of $r_{\text{rind}}$ can be written in terms of time ($t$) or in proper time ($\tau$). These definitions are mathematically equal.}.
From now on, we use $\alpha=a/c^2$ for brevity.
The coordinate transformation of Eq. (\ref{eq:Rtrans}) turns Eq. (\ref{eq:eom2}) into

\begin{equation}\label{eq:eom3}
    \frac{1}{c^2}\partial_\tau^2\varphi-\partial_\xi^2\varphi+e^{2\alpha\xi}\mu^2\varphi=e^{2\alpha\xi}j(\xi,\tau).
\end{equation}

\noindent The accelerated modes that solve the homogeneous differential equation yield the following expression,

\begin{equation}
    \psi_{\omega, \omega_0}(\xi,\tau)=\left(\frac{2\omega\sinh\pi\omega/\alpha c}{\alpha c\pi^2}\right)^{1/2}K_{i\omega/c \alpha}\left(\frac{\mu}{\alpha}e^{\alpha \xi}\right)e^{-i\omega_0 \tau},
\end{equation}

\noindent with $K$ the modified Bessel function of the second kind.
The accelerated modes satisfy the completeness relation \cite{higuchi_fulling-davies-unruh_1993}, which allows us to express the solution for a moving domain wall in terms of accelerated modes.
This is generally known as a Kontorovich–Lebedev transform \cite{Yakubovich1994} and it is given by,

\begin{equation}\label{eq:phiR}
    \varphi(\xi,\tau)=\int d\omega d\omega_0\frac{c J_R(\omega,\omega_0)\psi_{\omega, \omega_0}(\xi,\tau)}{\omega^2-\omega_0^2-i\varepsilon\omega_0},
\end{equation}

\noindent with 

\begin{equation}\label{eq:defJR}
    J_R(\omega,\omega_0)=\int d\xi d\tau \hspace{3px} j(\xi,\tau)\psi_{\omega, \omega_0}^*(\xi,\tau)e^{2\alpha \xi}.
\end{equation}

Using this result, we express $J_M$ in terms of $J_R$, which allows us to evaluate the average spin wave amplitude (Eq. \ref{eq:final}) for the moving domain wall.
This derivation is performed in detail in appendix \ref{Ap:A} and \ref{Ap:AM}.
We find that the average spin wave amplitude is given by the expression,

\begin{align}\label{eq:result}
    A_M&\approx \frac{16c^2\pi^5}{\mu V} \int \frac{d\omega}{\omega}|J_R(\omega,\omega)|^2\coth\left(\frac{\pi \omega}{\alpha c}\right). 
\end{align}

\noindent In this derivation, we assume that the proper acceleration $a$ is small compared to $c^2\mu/2$.
We find that the source $J_R$ is multiplied by $\coth\left(\pi \omega/\alpha c\right)=1+2n_B(2\pi \omega/\alpha c)$, where $n_B$ is the Bose-Einstein distribution at the Unruh temperature, $T_U=\hbar a / 2\pi c k_B$.
The origin of this factor is due to the difference between the stationary modes and the accelerated modes.
In particular, the accelerated modes are a mix of positive and negative frequency of stationary modes, $\coth\pi \omega/\alpha c=-n_B\left(-2\pi \omega/\alpha c\right)+n_B\left(2\pi \omega/\alpha c\right)$.
A more detailed explanation of this factor can be found in Ref. \cite{higuchi_fulling-davies-unruh_1993}.
The Bose-Einstein distribution at the Unruh temperature suggests that the excited spin waves due to the moving domain wall are thermal.
We discuss this further in Sec. \ref{sec:temp}.

\begin{figure}
    \centering
    \includegraphics[width=\linewidth]{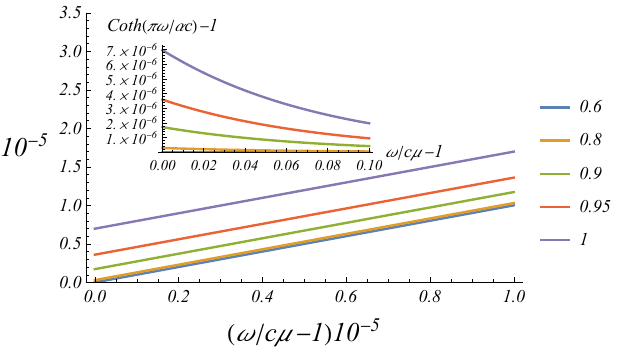}
    \caption{This figure shows $\frac{\omega}{c \mu}\coth\left(\pi \omega/\alpha c\right)-1$ as a function of frequency. The smaller figure shows $\coth\left(\pi \omega/\alpha c\right)-1=2n_B(2\pi \omega/\alpha c)$, which is the result of the Unruh effect. The legend refers to different values for acceleration, $2\alpha/\mu=2a/c^2 \mu$.}
    \label{fig:AmplificationFactor}
\end{figure}

It is important to compare the average spin wave amplitude of the stationary domain wall to that of the moving domain wall, since the difference between Eq. (\ref{eq:final}) and Eq. (\ref{eq:result}) is the dimensionless factor $\frac{\omega}{c \mu}\coth\left(\pi \omega/\alpha c\right)$, which is shown in Fig. \ref{fig:AmplificationFactor}.
Here, we see that for a moving domain wall, the average spin wave amplitude increases for larger accelerations.
In Fig. \ref{fig:AmplificationFactor}, we used a domain wall acceleration that is similar to $c^2\mu/2$.
However, when the domain wall accelerates faster, the approximation that we used to derive the average spin wave amplitude is no longer valid.
Fortunately, this amplitude can also be evaluated exactly when we use the domain wall ansatz to compute $j(x,t)=M_s^2(\gamma^2\eta H+i\gamma \dot \eta)/J A\gamma^2$. 
The domain wall ansatz gives $\eta=J'\sech((x-r_{\text{rind}}(t))/\lambda)e^{i\phi(t)}$.
In addition, we consider the domain wall to be accelerating close to $c$ and its width is small, $\lambda\ll e^{\alpha\xi_0}/\alpha$, such that we have

\begin{equation}
    j(\xi,\tau)\approx\pi\lambda \sigma e^{i\phi_0}\delta(\xi-\xi_0),
\end{equation}

\noindent with $\sigma=(J'M_s^2/JA)(H-c/\lambda\gamma)$, and $\phi_0=\hbar c (\beta-\alpha_G)/\beta \lambda K_\perp^{FM}$.
The explicit expression of $j(\xi,\tau)$ allows us to solve for $J_R$, which yields the expression,

\begin{equation}
    J_R(\omega,\omega_0)= \pi \lambda \sigma \delta(\omega_0) \psi^*_{\omega,\omega_0}(\xi_0,0)e^{2\alpha \xi_0+i\phi_0}.
\end{equation}

\noindent As before, we use $J_R$ to compute $J_M$ and $A_M$.
This is done by substituting our expression of $J_R$ into Eq. (\ref{eq:simp}), which solves Eq. (\ref{eq:final}).
We find that,

\begin{equation}
    A_M=\frac{32\pi^5\sigma^2\lambda^2}{\alpha^2\mu V}\left|K_0\left(\frac{\mu}{\alpha}e^{\alpha \xi_0}\right)\right|^2e^{4\alpha \xi_0}.
\end{equation}

\begin{figure}
    \centering
    \includegraphics[width=\linewidth]{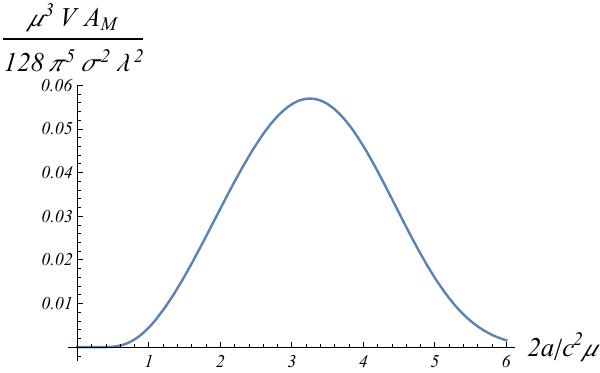}
    \caption{This figure shows the average spin wave amplitude excited by a source generated from an accelerated thin domain wall. The acceleration is made unitless using $2\alpha/\mu=2 a/c^2\mu$ and $\mu\xi_0$ is taken to be $1$.}
    \label{fig:AmforDWFigure}
\end{figure}

\noindent The average amplitude of spin waves generated by an accelerated domain wall is shown in Fig. \ref{fig:AmforDWFigure}.
We see that initially, the acceleration increases the amplitude.
However, at some point, the average amplitude reaches a maximum after which increasing the acceleration decreases the average amplitude.
This is because spin waves are less excited for faster accelerating domain walls, $\psi_{\omega,\omega_0}\propto 1/\sqrt{\alpha}$.

\subsection{Does the Unruh effect actually give temperature?}\label{sec:temp}

In this section, we briefly discuss the similarity of the Unruh effect and a thermal ensemble in a classical theory.
We start by deriving the spin wave amplitude for thermal fluctuations, after which we will point out the differences and similarities.

The stochastic field $\vec \zeta(x,t)$ emulates the thermal fluctuations.
The source in Eq. (\ref{eq:eom1}) is then changed to $\eta(x,t)=\zeta_x(x,t)+i\zeta_y(x,t)$ and its Fourier transform $\tilde \eta(k,\omega)$ satisfies \cite{stochfield}

\begin{equation}
    \langle \tilde \eta^*(k,\omega)\tilde \eta(k',\omega')\rangle\propto \frac{\omega}{\tanh(\omega/2k_BT)}\delta(k-k')\delta(\omega-\omega')
\end{equation}

\noindent with $T$ the temperature. 
The stochastic field is then used to compute the spin wave amplitude from Eq. (\ref{eq:final}),

\begin{equation}\label{eq:thermal}
    A_M\propto \int \frac{dk }{\omega(k)}\coth\left(\frac{\hbar\omega(k)}{2k_BT}\right).
\end{equation}

\noindent Comparing this result to Eq. (\ref{eq:result}) shows us a very similar form, with the crucial difference that the domain wall source is an additional factor in the integral.
In the case of the classical Unruh effect, the source continuously emits spin waves in a deterministic fashion, whereas thermal spin waves are generated stochastically.
This distinction implies that, when averaged over time, the classical Unruh effect produces a response that resembles thermal ensemble.
However, without averaging, the two processes remain fundamentally different.

In a quantum theory, this question becomes more nuanced because the state of the accelerating source appears as a mixed state when observed from a stationary detector.
Consequently, the mixed state is perceived as a thermal state characterised by the Unruh temperature, which is indistinguishable from a thermal ensemble.
While this correspondence strongly suggests a thermal interpretation, the question of whether the Unruh effect is truly thermal, in the strict thermodynamic sense, lies beyond the scope of this article.

\section{Discussion and Conclusion}\label{sec:Conclusion}

In this article, we demonstrated that a wire composed of a conducting ferromagnet exchange coupled to an insulating antiferromagnet, driven by a time-dependent charge current, can be used to simulate a magnetic analogue of the Unruh effect.
A domain wall in the conducting ferromagnet is a source of spin waves in the antiferromagnet, which is moved using a charge current in the wire.
When the domain wall is moved according to a Rindler trajectory, which is a trajectory of constant acceleration set by the spin waves in the antiferromagnet, the emission of spin waves in the antiferromagnet is characterised by the Unruh temperature, $T_U=\hbar a / 2\pi c k_B$.
This is the manifestation of the Unruh effect.

In our derivation of the Unruh effect, we have not considered damping effects on the spin waves in the antiferromagnet.
Spin wave damping is modelled by a finite $\varepsilon$ in Eq. (\ref{eq:phiR}).
If taken into account, the solution $\varphi$ has an additional factor given by $\exp(-\varepsilon^2 t/2)$.
This additional term modifies $A_M$ such that the factor $\exp(-\varepsilon^2 t)$ appears inside the time integral.
This factor lowers the average spin wave amplitude, but this does not change the results qualitatively.

Despite the reduction in average spin wave amplitude due to damping, the Unruh effect with $T_U=1K$ can be achieved when the acceleration of the ferromagnetic domain wall is equal to $a=2\pi c k_B T_U/\hbar$.
For the antiferromagnet DyFeO3, the analogue of the speed of light is approximately $20$km/s \cite{Hortensius2021}.
Thus, the acceleration of the ferromagnetic domain wall should be approximately $1.6\cdot 10^{16}$m/s${}^2$.
Ferromagnetic domain walls have been accelerated up to $10^{11}-10^{12}$m/s${}^2$, which could be even higher by considering ferromagnetic heterostructures or spin-orbit torque instead of spin transfer torque \cite{DW:mi15060696,DW:Yang2015}.
Nonetheless, the discrepancy may be too large to bridge.
To cross this gap, we need to lower the analogue of the speed of light in the antiferromagnet and increase the acceleration of the ferromagnetic domain wall.
Lower antiferromagnetic spin wave velocities may be achieved by decreasing the intersublattice antiferromagnetic exchange, such as in a synthetic antiferromagnet. 
The spin wave dispersion in the synthetic antiferromagnet in Ref. \cite{dispofSAFM} suggests that the maximum velocity is around $400$m/s.
For this velocity, the required acceleration of the domain wall reduces dramatically to $3.2\cdot 10^{14}$m/s$^2$.
The maximum group velocity could be reduced even more by using larger spacings in the synthetic antiferromagnet.
Optimising both the spin wave group velocity and the domain wall acceleration would allow for experimental verification of the effects we propose.

Given the experimental feasibility of observing the Unruh effect in this magnetic setup, it also offers a unique opportunity to explore its potential for generating and controlling quantum entanglement—a cornerstone of quantum technologies.
Previous theoretical studies have demonstrated that an accelerating atom emits entangled photon pairs \cite{scully_entanglement_2022}, while two co-accelerating two-level atoms become entangled \cite{benatti_entanglement_2004}. 
Additionally, the anti-Unruh effect that arises from finite-time and finite-space measurements \cite{brenna_anti-unruh_2016,garay_thermalization_2016} may produce even stronger entanglement than the conventional Unruh effect \cite{li_would_2018,pan_influence_2020,wu_does_2024,wu_genuine_2022}.
These findings suggest that similar entanglement dynamics could emerge in our system. 
Specifically, we anticipate that relativistic effects will facilitate entanglement between domain walls, particularly when two domain walls co-accelerate with an underlying antiferromagnetic layer. 
In this scenario, their internal angles are expected to become entangled, a phenomenon that has previously been studied using alternative entanglement strategies \cite{DWentanglement}.
Beyond domain walls, skyrmions have also been explored for quantum information applications \cite{skyr:Psaroudaki_Panagopoulos_2021}. 
Like domain walls, skyrmions may become entangled when co-accelerating in the proposed setup, as the analogue Unruh effect we propose here is independent of the specific magnetic texture.
This opens up a novel avenue for information manipulation through relativistic effects, aligning with the emerging field of quantum magnonics \cite{Yuan_Cao_Kamra_Duine_Yan_2022}.

Taken together, these findings demonstrate that antiferromagnets, with their inherent relativistic dynamics, offer a powerful platform not only for exploring high-energy phenomena but also for bridging the gap between fundamental physics and practical applications.

\begin{acknowledgments}
This work is funded by the projects “Black holes on a chip” with project number OCENW.KLEIN.502 and “Fluid Spintronics” with project number VI.C.182.069.
Both are financed by the Dutch Research Council (NWO).
\end{acknowledgments}

\appendix

\begin{widetext}

\section{Linearised antiferromagnetic spin wave theory with a local magnetic field}\label{Ap:Li}

In this appendix, we derive the equations of motion of long-wavelength spin waves on top of a two-sublattice antiferromagnet with uni-axial easy-axis anisotropy influenced by a magnetic field and a weakly coupled ferromagnetic layer.
The Lagrangian after integrating out the slave variable $\vec m=\vec m_1+\vec m_2$ is given by

\begin{align}
    \mathcal{L}[\vec n]=&\frac{M_s^2}{2J\gamma^2}\left|\dot{\vec n}-\gamma (\vec H+J'\vec n_{FM})\times \vec n\right|^2-\half(A|\partial_x\vec n|^2-Kn_z^2), \\
    =&\half\left(\frac{M_s^2}{J\gamma^2}|\dot{\vec n}|^2-A|\partial_x\vec n|^2+Kn_z^2\right)-\frac{M_s^2}{J\gamma}\dot{\vec n}\cdot \left((\vec H+J'\vec n_{FM})\times \vec n\right)+\frac{M_s^2}{2J}\left|(\vec H+J'\vec n_{FM})\times \vec n\right|^2.\label{Ap:L}
\end{align}

\noindent We linearise the Lagrangian using $\vec n=\hat z+\delta \vec n$ and $\delta \vec n=(\delta n_x,\delta n_y,\half\delta n_x^2 + \half\delta n_y^2)$.
In addition, we introduce the complex field $\varphi=\delta n_x+i\delta n_y$.
After linearising and introducing the complex field, the second term in Eq. (\ref{Ap:L}) is given by

\begin{align}
    \dot{\vec n}\cdot \left((\vec H+J'\vec n_{FM})\times \vec n\right)\approx & \delta\dot n_x\left(J'(\vec n_{FM})_y-(H+J'(\vec n_{FM})_z)\delta n_y\right)-\delta\dot n_y\left(J'(\vec n_{FM})_x-(H+J'(\vec n_{FM})_z)\delta n_y\right),\\
    =&(H+J'(\vec n_{FM})_z)(\delta\dot n_y\delta n_x-\delta\dot n_x\delta n_y)+J'(\vec n_{FM})_y\delta\dot n_x-J'(\vec n_{FM})_x\delta\dot n_y, \\
    =& \frac{1}{2i}(H+J'(\vec n_{FM})_z)(\dot \varphi \varphi^*- \varphi \dot\varphi^*)+i \half \eta^*\dot \varphi-i \half \eta \dot \varphi^*, \\
    \approx & -i H\dot \varphi \varphi^*-i \half\dot \eta^* \varphi+i \half\dot \eta \varphi^*.
\end{align}

\noindent Here, $\eta=J'(\vec n_{FM})_x+iJ'(\vec n_{FM})_y$, this is our spin wave source.
The same procedure is done for the third term in addition to omitting quadratic terms of $J' \vec n_{FM}$, since we assume that the coupling $J'$ is weak.

\begin{align}
    \left|(\vec H+J'\vec n_{FM})\times \vec n\right|^2\approx & H^2 \left(\delta n_y^2+\delta n_x^2\right)-2H\delta n_y(J'(\vec n_{FM})_y-J'(\vec n_{FM})_z \delta n_y)-2H\delta n_x(J'(\vec n_{FM})_x-J'(\vec n_{FM})_z \delta n_x), \\
    =& H(H+2J'(\vec n_{FM})_z)|\varphi|^2-\eta^* H\varphi-\eta H\varphi^*, \\
    \approx & H^2|\varphi|^2-\eta^* H\varphi-\eta H\varphi^*.
\end{align}

\noindent The linearised terms substituted back into the Lagrangian gives us

\begin{align}
    \mathcal{L}[\vec n]=&\half\left(\frac{M_s^2}{J\gamma^2}|\dot \varphi|^2-A|\partial_x \varphi|^2+K|\varphi|^2\right)-i\frac{M_s^2}{J\gamma}  H\dot \varphi \varphi^*+\half \frac{M_s^2}{J}H^2|\varphi|^2 \\
    &-\half\underbrace{\frac{M_s^2}{J\gamma^2}(\gamma^2\eta^* H-i\gamma \dot \eta^*)}_{A j^*} \varphi-\half\underbrace{\frac{M_s^2}{J\gamma^2}(\gamma^2\eta H+i\gamma \dot \eta)}_{A j}\varphi^*.\nonumber
\end{align}

\noindent This is the Lagrangian for antiferromagnetic spin waves with a source $Aj$.

\section{Computing the relation between $J_M$ and $J_R$}\label{Ap:A}

In this appendix, we express  $J_M$ in terms of $J_R$, which we will use to compute the average spin wave amplitude.
First, we recognise that the accelerated modes satisfy the completeness relation \cite{higuchi_fulling-davies-unruh_1993}

\begin{equation}
    \int d\xi d\tau\hspace{3px} \psi_{\omega, \omega_0}(\xi,\tau)\psi_{\omega', \omega_0'}(\xi,\tau)=c \delta(\omega-\omega')\delta(\omega_0-\omega_0'). 
\end{equation}

\noindent This allows us to invert Eq. (\ref{eq:defJR}),

\begin{equation}
    j(\xi,\tau)=e^{-2\alpha \xi}\int d\omega d\omega_0 \hspace{3px} J_R(\omega,\omega_0)\psi_{\omega, \omega_0}(\xi,\tau).
\end{equation}

\noindent The moving domain wall source can then be substituted into Eq. (\ref{eq:Jm}).
This allows us to express $J_M$ in terms of $J_R$ through

\begin{equation}
    J_M(k,\omega_0)=\int dxdt  \int d\omega' d\omega_0' \hspace{3px} J_R(\omega',\omega_0')\psi_{\omega', \omega_0'}(\xi,\tau)e^{-i kx+i\omega_0t-2\alpha \xi}.
\end{equation}

\noindent Now, we simplify $J_M$ by performing the transformation given in Eq. (\ref{eq:Rtrans}) and the change of variables $\eta=e^{\alpha c\tau}$,

\begin{align}
    J_M(k,\omega_0)=&  \int d\omega' d\omega_0' \hspace{3px} J_R(\omega',\omega_0')\int d\xi d\eta \hspace{3px} \left(\frac{2\omega'\sinh\pi\omega'/\alpha c}{\alpha c\pi^2}\right)^{1/2}K_{i\omega'/c \alpha}\left(\frac{\mu}{\alpha}e^{\alpha \xi}\right) \\
    &\times\frac{1}{\alpha c \eta}\exp\left[-i \frac{k}{\alpha}e^{\alpha \xi}\left(\eta+\frac{1}{\eta}\right)+i\frac{\omega_0}{\alpha c}e^{\alpha \xi}\left(\eta-\frac{1}{\eta}\right)-i\frac{\omega_0'}{\alpha c}\log(\eta)\right], \nonumber\\
    =&  \int d\omega' d\omega_0' \hspace{3px} J_R(\omega',\omega_0')\int d\xi d\eta \hspace{3px} \left(\frac{2\omega'\sinh\pi\omega'/\alpha c}{\alpha c\pi^2}\right)^{1/2}K_{i\omega'/c \alpha}\left(\frac{\mu}{\alpha}e^{\alpha \xi}\right) \\
    &\times\frac{\eta^{-i\frac{\omega_0'}{\alpha c}-1}}{\alpha c}\exp\left[\frac{ie^{\alpha \xi}}{2\alpha c}\left(\omega_0-ck)\right)\left(\eta-\frac{1}{\eta}\left(\frac{\omega_0+ck}{\omega_0-ck}\right)\right)\right].\nonumber
\end{align}

\noindent The last term is simplified by using the following identity (Eq. (3.471.10) in Ref. \cite{Integral}),

\begin{equation}
    \int d\eta \hspace{3px} \eta^{\nu-1}\exp\left[\frac{iu}{2}\left(\eta-\frac{\beta^2}{\eta}\right)\right]=2\beta^\nu e^{i\pi\nu/2}K_{-\nu}(\beta u).
\end{equation}

\noindent The expression of $J_M$ is then given by

\begin{align}
    J_M(k,\omega_0)=&  \int d\omega' d\omega_0' \hspace{3px} J_R(\omega',\omega_0')\int d\xi \hspace{3px} \left(\frac{2\omega'\sinh\pi\omega'/\alpha c}{\alpha c\pi^2}\right)^{1/2}K_{i\omega'/c \alpha}\left(\frac{\mu}{\alpha}e^{\alpha \xi}\right) \\
    &\times\frac{2}{\alpha c}\left(\frac{\omega_0+ck}{\omega_0-ck}\right)^{-i\omega_0'/2\alpha c}e^{\pi \omega_0'/2 \alpha c}K_{i\omega_0'/c \alpha}\left(\frac{(\omega_0^2-c^2k^2)^{1/2}}{\alpha c}e^{\alpha \xi}\right).\nonumber
\end{align}

\noindent For the calculation hereafter, it is necessary to set $\omega_0=\pm \omega(k)$, which simplifies the expression of $J_M$ further,

\begin{align}
    J_M(k,\pm\omega(k))=&  \int d\omega' d\omega_0' \hspace{3px} J_R(\omega',\omega_0')  \frac{2}{\alpha c}\left(\frac{\omega(k)+ck}{\omega(k)-ck}\right)^{\mp i\omega_0'/2\alpha c}e^{\pi \omega_0'/2 \alpha c} \\
    &\times\left(\frac{2\omega'\sinh\pi\omega'/\alpha c}{\alpha c\pi^2}\right)^{1/2}\int d\xi \hspace{3px} K_{i\omega_0'/c \alpha}\left(\frac{\mu}{\alpha }e^{\alpha \xi}\right)K_{i\omega'/c \alpha}\left(\frac{\mu}{\alpha}e^{\alpha \xi}\right). \nonumber
\end{align}

\noindent Using the completeness relation of the modified Bessel function of the second kind gives us the following expression

\begin{align}
    J_M(k,\pm\omega(k))=&  \int d\omega' \hspace{3px} J_R(\omega',\omega')  \frac{2}{\alpha c}\left(\frac{\omega(k)+ck}{\omega(k)-ck}\right)^{\mp i\omega'/2\alpha c}\left(\frac{\alpha c\pi^2}{2\omega'\sinh\pi\omega'/\alpha c}\right)^{1/2}e^{\pi \omega'/2 \alpha c} \\
    &+ \int d\omega' \hspace{3px} J_R(\omega',-\omega')  \frac{2}{\alpha c}\left(\frac{\omega(k)+ck}{\omega(k)-ck}\right)^{\pm i\omega'/2\alpha c}\left(\frac{\alpha c\pi^2}{2\omega'\sinh\pi\omega'/\alpha c}\right)^{1/2}e^{-\pi \omega'/2 \alpha c}.\nonumber
\end{align}

\noindent This result allows us to compute the average spin wave amplitude for an accelerating source in the following appendix.

\section{Computing $A_M$}\label{Ap:AM}

Herein, we use the result of appendix \ref{Ap:A} to compute the average spin wave amplitude.
First, we define

\begin{align}
    g(\omega')&=\frac{2}{\alpha c}\left(\frac{\alpha c\pi^2}{2\omega'\sinh\pi\omega'/\alpha c}\right)^{1/2}e^{\pi \omega'/2 \alpha c}. 
\end{align}

\noindent Now, we solve for $A_M$ by substituting our expression of $J_M$,

\begin{align}\label{eq:link2}
    \int \frac{dk}{\omega^2(k)}|J_M(\omega,\pm\omega)|^2= &\int\frac{dk}{\omega^2(k)} \int d\omega'd\omega''\bigg[J_R(\omega',\omega')J_R^*(\omega'',\omega'')g(\omega')g(\omega'')\left(\frac{\omega(k)+ck}{\omega(k)-ck}\right)^{\mp i(\omega'-\omega'')/2\alpha c}\\
    &+J_R(\omega',-\omega')J_R^*(\omega'',-\omega'')g(-\omega')g(-\omega'')\left(\frac{\omega(k)+ck}{\omega(k)-ck}\right)^{\pm i(\omega'-\omega'')/2\alpha c} \nonumber\\
    &+J_R(\omega',\omega')J_R^*(\omega'',-\omega'')g(\omega')g(-\omega'')\left(\frac{\omega(k)+ck}{\omega(k)-ck}\right)^{\mp i(\omega'+\omega'')/2\alpha c} \nonumber\\
    &+J_R(\omega',-\omega')J_R^*(\omega'',\omega'')g(-\omega')g(\omega'')\left(\frac{\omega(k)+ck}{\omega(k)-ck}\right)^{\pm i(\omega'+\omega'')/2\alpha c} \bigg]. \nonumber
\end{align}

\noindent The $k$ integral is solved by performing the transformation $v=\ln\left[(\omega(k)+ck)/(\omega(k)-ck)\right]$, which yields the expression,

\begin{align}
    \int_{-\infty}^\infty  \frac{dk}{\omega^2(k)}\left(\frac{\omega(k)+ck}{\omega(k)-ck}\right)^{-i(\omega'-\omega'')/2\alpha c}=&\frac{1}{2c^2\mu}\int_{-\infty}^\infty \frac{dv}{\cosh(v/2)}e^{-iv(\omega'-\omega'')/2\alpha c}, \\
    =&\frac{\pi}{c^2\mu}\sech\left(\frac{\pi(\omega'-\omega'')}{2\alpha c}\right).
\end{align}

\noindent This result is substituted back into Eq. (\ref{eq:link2}), which gives us the following,

\begin{align}
    \int  \frac{dk}{\omega^2(k)}|J_M(k,\pm\omega(k))|^2= & \frac{\pi}{c^2\mu} \int d\omega'd\omega''\big[J_R(\omega',\omega')J_R^*(\omega'',\omega'')g(\omega')g(\omega'')\sech\left(\frac{\pi(\omega'-\omega'')}{2\alpha c}\right) \label{eq:simp}\\
    &+J_R(\omega',-\omega')J_R^*(\omega'',-\omega'')g(-\omega')g(-\omega'')\sech\left(\frac{\pi(\omega'-\omega'')}{2\alpha c}\right) \nonumber\\
    &+J_R(\omega',\omega')J_R^*(\omega'',-\omega'')g(\omega')g(-\omega'')\sech\left(\frac{\pi(\omega'+\omega'')}{2\alpha c}\right) \nonumber\\
    &+J_R(\omega',-\omega')J_R^*(\omega'',\omega'')g(-\omega')g(\omega'')\sech\left(\frac{\pi(\omega'+\omega'')}{2\alpha c}\right) \big], \nonumber\\
    \approx & \frac{2\alpha \pi}{c\mu} \int d\omega'\big[|J_R(\omega',\omega')|^2g(\omega')^2+|J_R(\omega',-\omega')|^2g(-\omega')^2\\
    &+J_R(\omega',\omega')J_R^*(-\omega',\omega')g(\omega')^2+J_R(\omega',-\omega')J_R^*(-\omega',-\omega')g(-\omega')^2 \big].\nonumber
\end{align}

\noindent In the last equality sign, we used that $\frac{\pi}{c^2\mu}\sech\left(\frac{\pi(\omega'-\omega'')}{2\alpha c}\right)\approx \frac{2\alpha \pi}{c \mu}\delta(\omega'-\omega'')$ when $2\alpha /\mu \ll 1$.
This is equivalent to a proper acceleration that is small, $a\ll c^2\mu/2$.
The approximate solution is then substituted back into Eq. (\ref{eq:final}).
The result is therefore given by

\begin{align}
    A_M&\approx \frac{4 \alpha c^3\pi^3}{\mu V} \int d\omega'\big[|J_R(\omega',\omega')|^2g(\omega')^2+|J_R(\omega',-\omega')|^2g(-\omega')^2\\
    &+J_R(\omega',\omega')J_R^*(-\omega',\omega')g(\omega')^2+J_R(\omega',-\omega')J_R^*(-\omega',-\omega')g(-\omega')^2 \big], \nonumber \\
    &= \frac{4\alpha c^3\pi^3}{\mu V} \int d\omega'|J_R(\omega',\omega')|^2(g(\omega')^2+g(-\omega')^2), \nonumber \\
    &=\frac{16c^2\pi^5}{\mu V} \int \frac{d\omega'}{\omega'}|J_R(\omega',\omega')|^2\coth\left(\frac{\pi \omega'}{\alpha c}\right).
\end{align}

\noindent Additionally, we used that $|J_R(\omega',\omega')|^2=|J_R(\omega',-\omega')|^2$, also $g(\omega')^2+g(-\omega')^2=\frac{4\pi^2}{\alpha c \omega}\coth\left(\frac{\pi \omega'}{\alpha c}\right)$.

\end{widetext}

\nocite{*}

\bibliography{export-data}

\end{document}